\documentclass[aps,prl,twocolumn,superscriptaddress]{revtex4-2}

\usepackage{amssymb}
\usepackage{amsmath}
\usepackage{graphicx}
\usepackage{array}
\usepackage{multirow}
\usepackage{setspace}
\usepackage[colorlinks,linkcolor=blue,anchorcolor=blue,citecolor=red]{hyperref} 

\begin{document}

\title{The electronic structures, magnetic transition and Fermi surface instability of room-temperature altermagnet KV$_{2}$Se$_{2}$O}
\author{Yuanji Xu}
\affiliation{Institute for Applied Physics and Department of Physics, University of Science and Technology Beijing, Beijing 100083, China}
\author{Huiyuan Zhang}
\affiliation{Institute for Applied Physics and Department of Physics, University of Science and Technology Beijing, Beijing 100083, China}
\author{Maoyuan Feng}
\affiliation{Institute for Applied Physics and Department of Physics, University of Science and Technology Beijing, Beijing 100083, China}
\author{Fuyang Tian}
\email{fuyang@ustb.edu.cn}
\affiliation{Institute for Applied Physics and Department of Physics, University of Science and Technology Beijing, Beijing 100083, China}
\date{\today}

\begin{abstract}
Altermagnetism has recently emerged as a distinct and fundamental class of magnetic order. Exploring its interplay with quantum phenomena such as unconventional superconductivity, density-wave instabilities and many-body effects represents a compelling frontier. In this work, we theoretically confirm the presence of high-temperature metallic altermagnetism in KV$_2$Se$_2$O. We demonstrate that the anomalous metal-insulator-metal transition arises from a Lifshitz transition associated with Fermi surface reconstruction. The previously reported spin-density wave gap is found to lie below the Fermi level in our study and is now recognized to be attributed to the V-shaped density of states, originating from orbital-selective and sublattice-resolved half-metal-like behavior on a specific V atom. Furthermore, we identify the instability from the nesting of spin-momentum-locked one-dimensional Fermi surfaces, which induces the SDW state. These findings position KV$_2$Se$_2$O as a promising platform for investigating the interplay among altermagnetism, unconventional superconductivity and density-wave order.
\end{abstract}

\maketitle

Altermagnetism represents a newly identified class of long-range magnetic order characterized by non-relativistic spin splitting in the absence of net magnetization \cite{Smejkal2022,Smejkal2022.2,Mazin2022}. In altermagnets, the two spin sublattices are related by point-group symmetries such as rotations or mirror operations within the framework of spin space groups, distinct from the inversion or translation symmetries that relate sublattices in conventional antiferromagnets \cite{Xiao2024,Chen2024,Jiang2024}. This symmetry-driven, momentum-dependent spin splitting gives rise to spin-momentum locking, offering significant advantages for spintronic applications and providing a fertile ground for exploring unconventional physical phenomena \cite{Jungwirth2016,Bai2024}. More broadly, magnetism plays a central role in condensed matter physics, underpinning a wide range of emergent states, including unconventional superconductivity  \cite{Scalapino2012}, metal-insulator transitions \cite{Lee2018}, Hund metals \cite{Yin2011}, and spin-density wave (SDW) orders \cite{Hu2022}. The discovery of altermagnetism thus opens a new frontier for investigating exotic quantum phenomena driven by this unconventional form of magnetic order \cite{Liu2024,Xu2025.alt}.

Although symmetry-based spin space group analyses predict a large number of potential altermagnetic materials, only a few candidates have been experimentally confirmed to host altermagnetism \cite{Song2025,Krempasky2024,Lee2024,Zhu2024,Reimers2024,Zhou2025}. Recently, KV$_2$Se$_2$O was identified as a room-temperature $d$-wave altermagnet through spin- and angle-resolved photoemission spectroscopy (SARPES) \cite{Jiang2025}. Unlike altermagnetic semiconductors, the metallic nature of KV$_2$Se$_2$O offers a rare opportunity to explore the interplay between unconventional altermagnetic order and low-energy quasiparticle phenomena, including superconductivity, density-wave (DW)-like behavior, and many-body effects. SARPES measurements reveal folded bands near the Fermi level and zero-field nuclear magnetic resonance (NMR) spectra show a splitting into two distinct sets of peaks, suggesting the emergence of SDW phase below 100 K. However, temperature-dependent X-ray diffraction and scanning tunneling microscopy detect no signs of a charge-density wave (CDW) or associated lattice modulation \cite{Jiang2025,Zhuang2025}. As a result, the microscopic nature of the DW-like phase transition remains unresolved.

In addition, transport measurements reveal an anomalous feature in the resistivity near the DW-like transition, characterized by a metal-to-insulator–like behavior \cite{Bai2024_2}. Similar resistive anomalies have been reported in related materials and are typically attributed to the partial gapping of the Fermi surface due to SDW or CDW formation \cite{Shi2013,Ozawa2000,Ablimit2018}. However, in KV$_2$Se$_2$O, Hall coefficient measurements indicate a substantial reduction in carrier density, by nearly an order of magnitude, between 300 K and 105 K, well above the DW-like transition temperature \cite{Bai2024_2}. Furthermore, angle-resolved photoemission spectroscopy (ARPES) measurements reveal pseudogap-like features, purportedly originating from SDW fluctuations, that persist up to 200 K, well above the SDW transition temperature \cite{Jiang2025}. These puzzling observations challenge the conventional interpretation of the DW-like transition in KV$_2$Se$_2$O and underscore the need for a more comprehensive theoretical investigation into its electronic structures and magnetism.

\begin{figure}
\begin{center}
\includegraphics[width=0.48\textwidth]{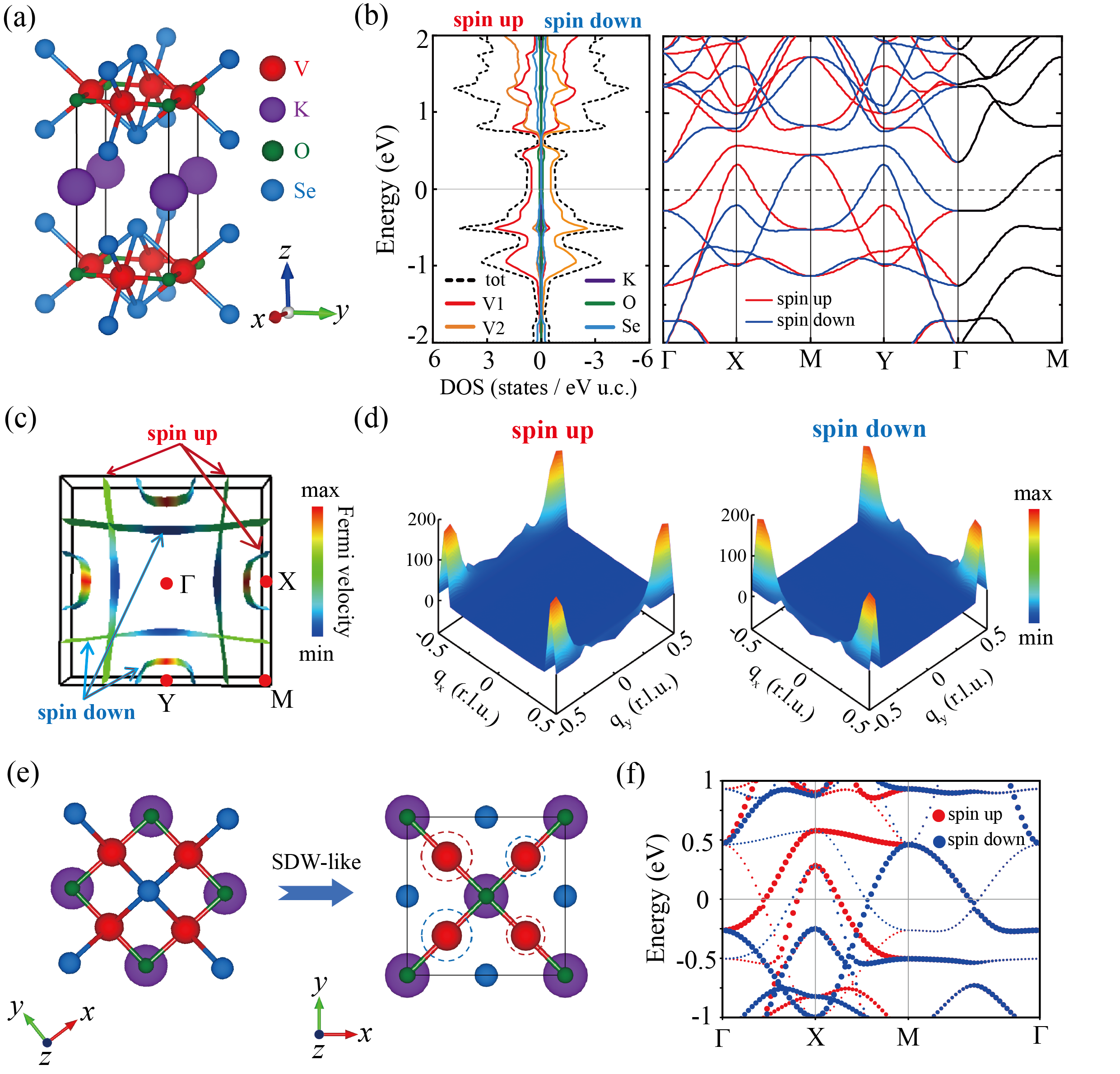}\caption{(a) Crystal structure of KV$_2$Se$_2$O. (b) Electronic band structure and projected density of states (PDOS) of the altermagnetic phase calculated within DFT. (c) Fermi surfaces and corresponding Fermi velocities of the altermagnetic phase. (d) Real part of the dynamical spin susceptibility in the zero-frequency limit, showing prominent nesting at $\boldsymbol{Q} = (0.5, 0.5, 0)$ in reciprocal lattice units (r.l.u.) for both spin-up and spin-down channels. (e) Top view of the KV$_2$Se$_2$O crystal structure and its magnetic configuration in the SDW phase. The size of the dashed circles represents the magnitude of the magnetic moments. (f) Unfolded band structure of the SDW phase obtained from DFT calculations.}
\label{fig1}
\end{center}
\end{figure}

In this work, we systematically investigate the electronic structure and magnetic transitions of KV$_2$Se$_2$O using density functional theory combined with dynamical mean-field theory (DFT+DMFT) \cite{Georges1996,Kotliar2006,Haule2010}. Our results reveal that the altermagnetic state remains robust well above room temperature, accompanied by a significantly reduced magnetic moment due to dynamical correlation effects, consistent with experimental observations. Notably, the symmetry of the altermagnetic order gives rise to a spin- and orbital-selective ``half-metallic” behavior on a specific vanadium atom, where electronic bands of only one spin channel cross the Fermi level while the other remains gapped. This asymmetry leads to a pseudogap feature in the total density of states (DOS), previously interpreted as an SDW gap. Moreover, the disappearance of electron-like Fermi pockets near the $\Gamma$ point around the DW-like transition temperature is indicative of a Lifshitz transition, offering a natural explanation for the observed metal-insulator-metal behavior. Finally, our momentum-resolved susceptibility calculations reveal a pronounced Fermi surface nesting instability, providing microscopic evidence for the emergence of a spin-density wave at low temperatures.

\begin{figure*}
\begin{center}
\includegraphics[width=0.73\textwidth]{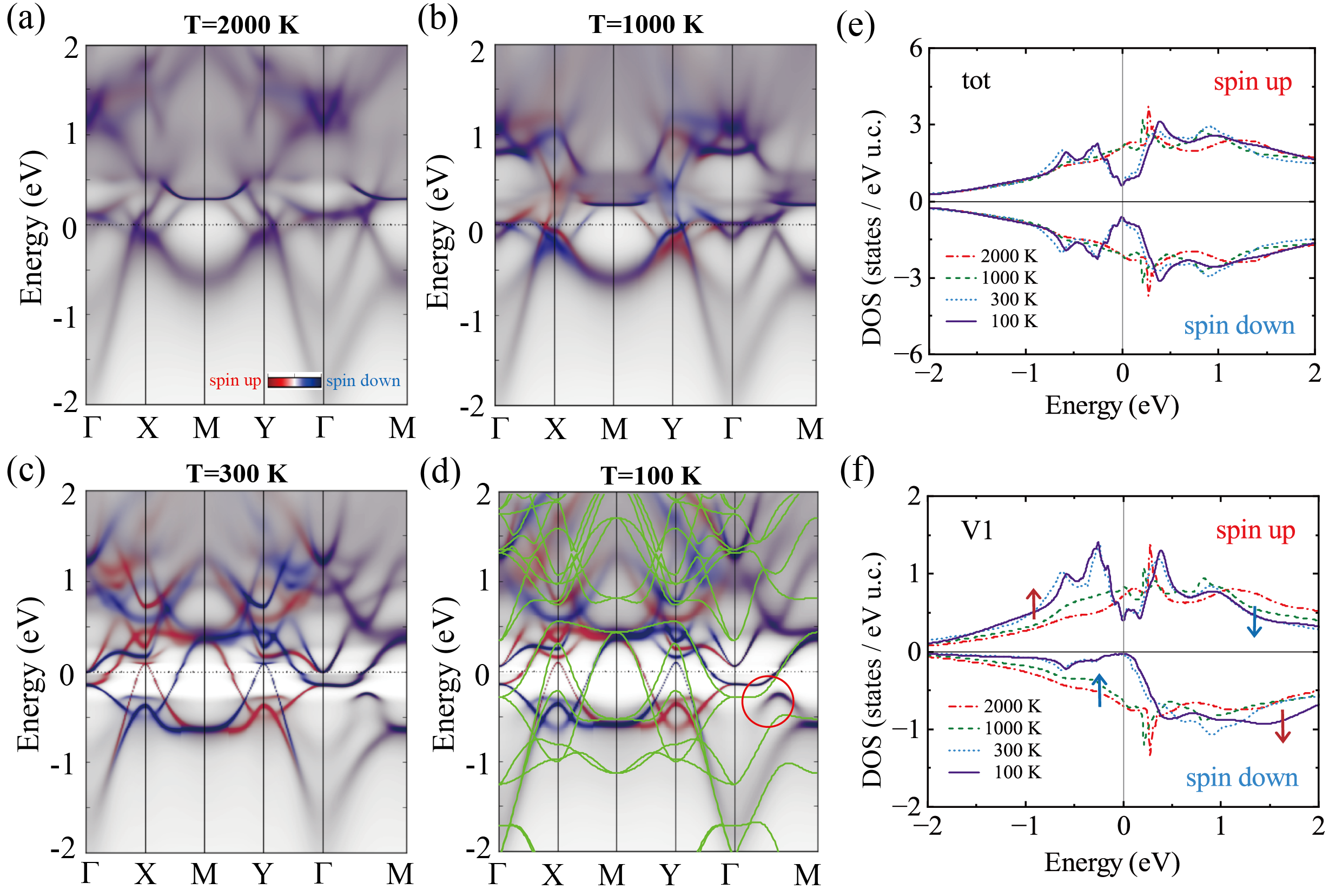}
\caption{Results of DFT+DMFT calculations for KV$_2$Se$_2$O at various temperatures. (a)–(d) Momentum-resolved spectral functions computed from 2000 K to 100 K. The green curves in (d) represent the corresponding DFT band structure for comparison. (e) Total density of states for spin-up and spin-down channels across the same temperature range. (f) The LDOS projected onto the V1 atom for both spin channels from 2000 K to 100 K. Blue arrows indicate areas where the DOS decreases with decreasing temperature, while red arrows indicate areas where the DOS increases with decreasing temperature.}
\label{fig2}
\end{center}
\end{figure*}

The oxychalcogenide KV$_2$Se$_2$O crystallizes in a tetragonal structure with an anti-K$_2$NiF$_4$-type layered configuration, as illustrated in Fig.~\ref{fig1}(a). It belongs to the $P4/mmm$ space group, where the V$_2$O planes form anti-CuO$_2$-like units, with two Se atoms positioned above and below the center of each V$_2$O square. Potassium atoms are intercalated between these layers along the $c$-axis. Structural relaxation, including both lattice constants and internal atomic positions, was performed using the VASP package with the projector augmented-wave method (see Supplementary Material) \cite{Kresse1996,Bloch1994}. The optimized lattice parameters in the altermagnetic phase are $a$ = 3.98\,\AA{} and $c$ = 7.48\,\AA{}, in agreement with experimental measurements \cite{Bai2024_2}. Electronic structure calculations were conducted using the full-potential linearized augmented plane-wave method, as implemented in the WIEN2k package \cite{wien2k2023}. The exchange-correlation potential was treated within the Perdew-Burke-Ernzerhof generalized gradient approximation \cite{Perdew1996}. As shown in Fig.~\ref{fig1}(b), our DFT results successfully reproduce the key features reported in previous studies \cite{Jiang2025}. In the altermagnetic configuration, the total DOS resembles a conventional antiferromagnet, with identical spin-up and spin-down DOS due to the absence of net magnetization. The electronic states near the Fermi level are dominated by V-$d$ electrons, indicating the itinerant nature of magnetism in this material. Notably, due to the spin space group symmetry [$C_2$\textbar\textbar$C_{4z}$], the spin-up and spin-down bands remain degenerate along the $\Gamma$–$M$ high-symmetry path, exhibiting a $d$-wave–like form, characteristic of altermagnetic order.

Besides, it is important to emphasize that in the spin-up channel, only the V1 atom exhibits a finite local DOS (LDOS) at the Fermi level, while the V2 atom displays a pronounced gap. Conversely, in the spin-down channel, the situation is reversed. Such atomic- and spin-selective behavior leads to a pronounced spin-resolved anisotropy in the Fermi surfaces, as shown in Fig.~\ref{fig1}(c). The flat, quasi-one-dimensional (1D) Fermi surface sheets in the spin-up channel extend along the $k_y$–$k_z$ plane and originate predominantly from the V1 atoms located at $(0.5,0,0)$, corresponding to spin-up current transport along the $x$-direction. Similarly, the quasi-1D Fermi surfaces in the spin-down channel are primarily derived from the V2 atoms at $(0,0.5,0)$, facilitating spin-down current along the $y$-direction. This spin-momentum-locked transport behavior reflects the strong V-O bonding and the underlying rotational symmetry inherent to the altermagnetic order. In addition, the geometry of the Fermi surfaces exhibits clear signatures of strong nesting. As shown in Fig.~\ref{fig1}(d), the calculated susceptibility of the Fermi surfaces reveals a pronounced nesting peak at the wave vector $\boldsymbol{Q} = (0.5, 0.5, 0)$ in reciprocal lattice units. This nesting vector is consistent with the SDW instability proposed by recent experimental observations \cite{Jiang2025}, as illustrated in Fig.~\ref{fig1}(e).

However, it is important to note that the unfolded band structure of the SDW phase resembles that of previous theoretical calculations, while the SDW gap is located below the Fermi level, as shown in Fig.~\ref{fig1}(f) \cite{Jiang2025}. Our results are consistent with recent theoretical work reporting an SDW gap approximately 0.15 eV below the Fermi energy \cite{Yan2025}. Furthermore, the magnetic moment obtained in our DFT calculations is approximately 1.7 $\mu_{B}$, significantly larger than the experimentally observed value of 0.7 $\mu_{B}$ extracted from NMR measurements \cite{Jiang2025}. This overestimation may influence the position of the reconstructed bands in the SDW state (see Supplementary Material). Therefore, the commonly assumed mechanism, that the metal-insulator-metal transition observed in resistivity arises from Fermi surface gapping due to the SDW transition, should be reconsidered with caution.

To further investigate the role of dynamical correlations, we performed fully charge self-consistent DFT+DMFT calculations on KV$_2$Se$_2$O to examine its temperature-dependent electronic structure and magnetism. The double-counting correction was treated using the exact double-counting scheme \cite{Haule2015}. The hybridization-expansion continuous-time quantum Monte Carlo method was used in the impurity solver \cite{Werner2006,Haule2007}. An Ising-type Hund’s coupling was adopted, with on-site Coulomb interaction parameters set to $U = 7.0$ eV and $J = 0.8$ eV (see Supplementary Material). At high temperatures, the system remains in a paramagnetic state, as shown in Fig.~\ref{fig2}(a). However, upon lowering the temperature, momentum-dependent spin splitting begins to emerge, becoming evident even at 1000 K, as illustrated in Fig.~\ref{fig2}(b). At these elevated temperatures, the spectral functions are broadly incoherent across a wide energy range, indicative of strong electronic scattering. This behavior is also reflected in the total DOS, which displays broad peaks at high temperatures in Fig.~\ref{fig2}(e). As the temperature is reduced to room temperature, sharp quasiparticle bands emerge near the Fermi level, while incoherent features persist at higher energies, as shown in Fig.~\ref{fig2}(c). At 100 K, as shown in  Fig.~\ref{fig2}(d), the quasiparticle features become even more well-defined, while the overall band structure remains qualitatively unchanged. The minimal temperature dependence of the Fermi surface is consistent with recent ARPES measurements \cite{Jiang2025}.

The emergence of quasiparticle coherence near the Fermi level, accompanied by the persistence of incoherent spectral weight at higher energies across the magnetic transition, prompts a reconsideration of the dynamical correlation model recently proposed by Xu et al. for itinerant ferromagnets \cite{Xu2024,Xu2025,Xu2025_2}. In altermagnetic systems, the hallmark asymmetric transfer of spectral weight between spin channels can be discerned through analysis of the LDOS on individual V-site. As illustrated in Fig.~\ref{fig2}(f), in the spin-up channel, spectral weight above the Fermi level decreases with decreasing temperature, while the weight below the Fermi level increases. Conversely, the spin-down channel exhibits the opposite trend. Furthermore, the calculated magnetic moment of V-atoms in the altermagnetic phase is approximately 0.8 $\mu_B$, significantly reduced compared to the DFT value and in excellent agreement with the NMR measurements \cite{Jiang2025}. This substantial suppression of the magnetic moment arises from dynamical correlation effects, further supporting the applicability of the dynamical correlation model to the itinerant altermagnet.

Our DFT+DMFT calculations reveals a pronounced V-shaped feature in the total DOS at the Fermi level in the altermagnetic phase in Fig.~\ref{fig2}(e). Moreover, as illustrated in Fig.~\ref{fig2}(f), the LDOS of the V1-atom displays a gap in the spin-down channel, while the spin-up channel retains finite spectral weight at the Fermi level. This spin-selective ``half-metallic” character on a specific V-site is consistent with our DFT results. The resulting  total DOS at the Fermi level is approximately 1.26 states/eV f.u., yielding a Sommerfeld coefficient  $\gamma_0 = 2.97\,$mJ/mol K$^2$, slightly larger than the experiment value of 1.9 \cite{Bai2024_2}. Our DFT+DMFT results further indicate that the V-atoms exhibit mixed-valence behavior, with an average V-$d$ occupancy varying from 2.6 to 2.7 electrons between 2000 K and 100 K. This reflects the itinerant character of the V-$d$ electrons and the presence of strong charge fluctuations. Notably, this temperature-dependent mixed-valence behavior and the increase in $d$-electron count with decreasing temperature are also supported by recent electron energy loss spectroscopy measurements \cite{Zhuang2025}.

\begin{figure}
\begin{center}
\includegraphics[width=0.45\textwidth]{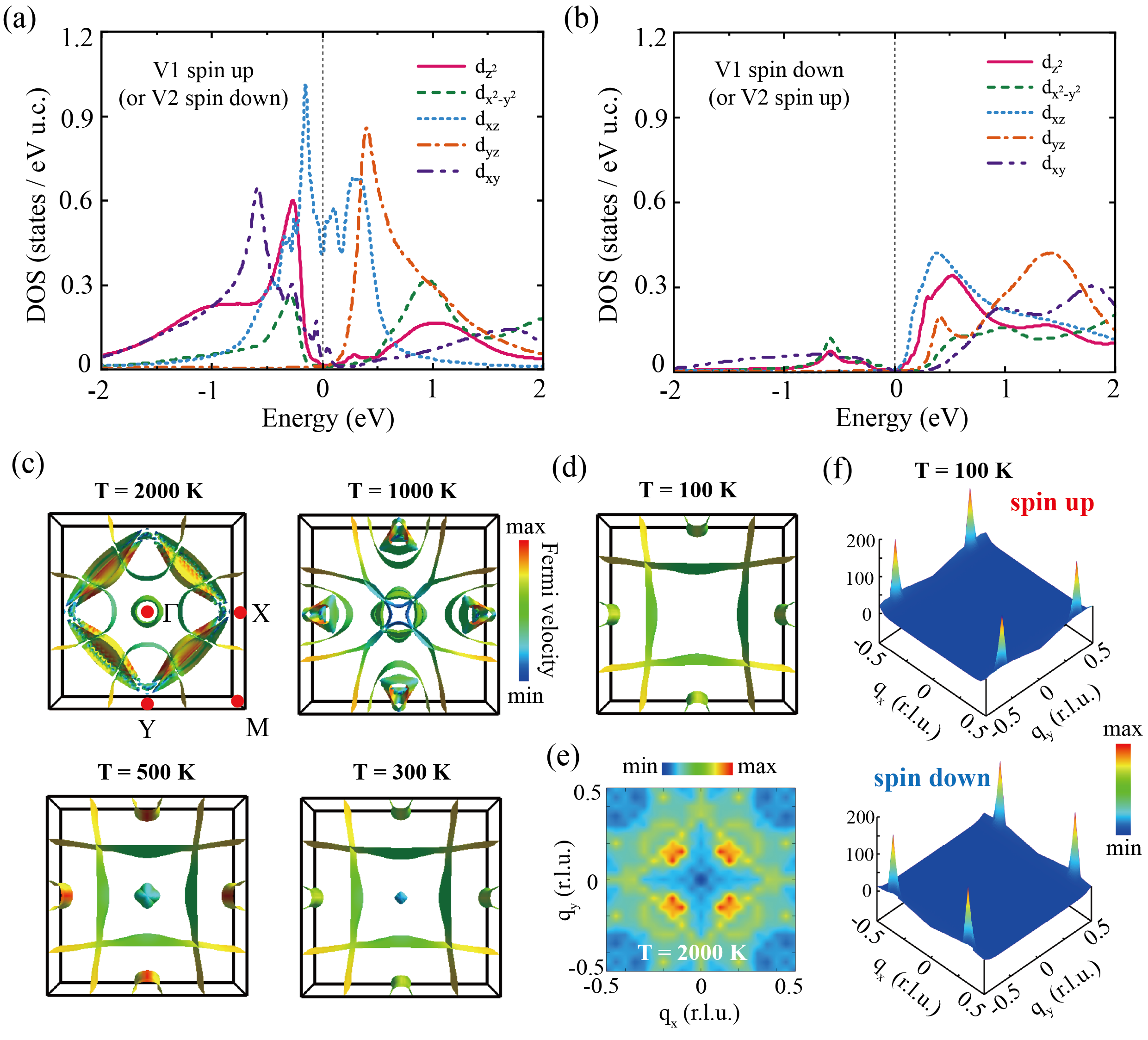}
\caption{(a) PDOS for V1 electrons in the spin-up channel (equivalently, V2 electrons in the spin-down channel). (b) PDOS for V1 electrons in the spin-down channel (equivalently, V2 electrons in the spin-up channel). The local axis of the V2 atom is rotated by ${90}^{\circ}$ relative to V1, in accordance with spin-group symmetry. (c) Fermi surfaces obtained from DFT+DMFT calculations at temperatures ranging from 2000 K to 300 K. (d) Fermi surfaces from DFT+DMFT calculations at 100 K. (e) Real part of the dynamical susceptibility in the zero-frequency limit at 2000 K (paramagnetic phase). (f) Real part of the dynamical susceptibility at zero frequency obtained from DFT+DMFT calculations at 100 K, showing pronounced nesting features at $\boldsymbol{Q} \approx (0.5,0.5,0)$ (in reciprocal lattice units) in both spin-up and spin-down channels.}
\label{fig3}
\end{center}
\end{figure}

Recent experimental studies have suggested that the anomalous metal-insulator-metal transition observed in the resistivity of KV$_2$Se$_2$O may be associated either with changes in the V-valence state or to the opening of the SDW gap at the Fermi level \cite{Jiang2025,Bai2024_2,Zhuang2025}. Instead, our DFT+DMFT calculations reveal a pronounced temperature-dependent shift of the electronic bands near the $\Gamma$ point. As shown in Fig.~\ref{fig2}, the gradual suppression and eventual disappearance of this electron-like Fermi pocket centered at the $\Gamma$ point correlates closely with the continuous decrease in carrier density and its sharp drop near 105 K, as observed in Hall measurements \cite{Bai2024_2}. As illustrated in Fig.~\ref{fig3}(c), the Fermi surface undergoes a substantial topological transformation across the altermagnetic phase transition. When entering the altermagnetic phase, electron-like Fermi pocket remains clearly visible at the $\Gamma$ point even at room temperature. However, it vanishes at around 100 K, as shown in Fig.~\ref{fig3}(d). This temperature-dependent reconstruction of the Fermi surface provides a compelling explanation for the observed metal-insulator behavior in resistivity, consistent with a Lifshitz transition scenario.

Moreover, as discussed earlier, the LDOS of individual V atom in our DFT+DMFT calculations reveals a spin-selective ``half-metallic" character and V-shape character. As illustrated in Fig.~\ref{fig3}(a) and (b), the PDOS at the Fermi level for V1 atom is dominated by the $d_{xz}$ orbital character. Similarly, the Fermi surfaces in the spin-down channel originate mainly from V2-$d_{yz}$ electrons (in the global coordinates). This orbital and spin-selective electronic structures provide a direct evidence for spin-momentum-locked current in the altermagnetic state. Finally, we evaluated the susceptibility based on the DFT+DMFT calculated Fermi surfaces. In the paramagnetic phase, the susceptibility does not exhibit a prominent nesting vector, as shown in Fig.~\ref{fig3}(e). However, in the altermagnetic phase, a pronounced nesting peak emerges at $\boldsymbol{Q} \approx (0.5, 0.5, 0)$, as shown in Fig.~\ref{fig3}(f). These results provide compelling evidence of Fermi surface instability and supports the formation of SDW phase at low temperatures in KV$_2$Se$_2$O.

\begin{figure}
\begin{center}
\includegraphics[width=0.45\textwidth]{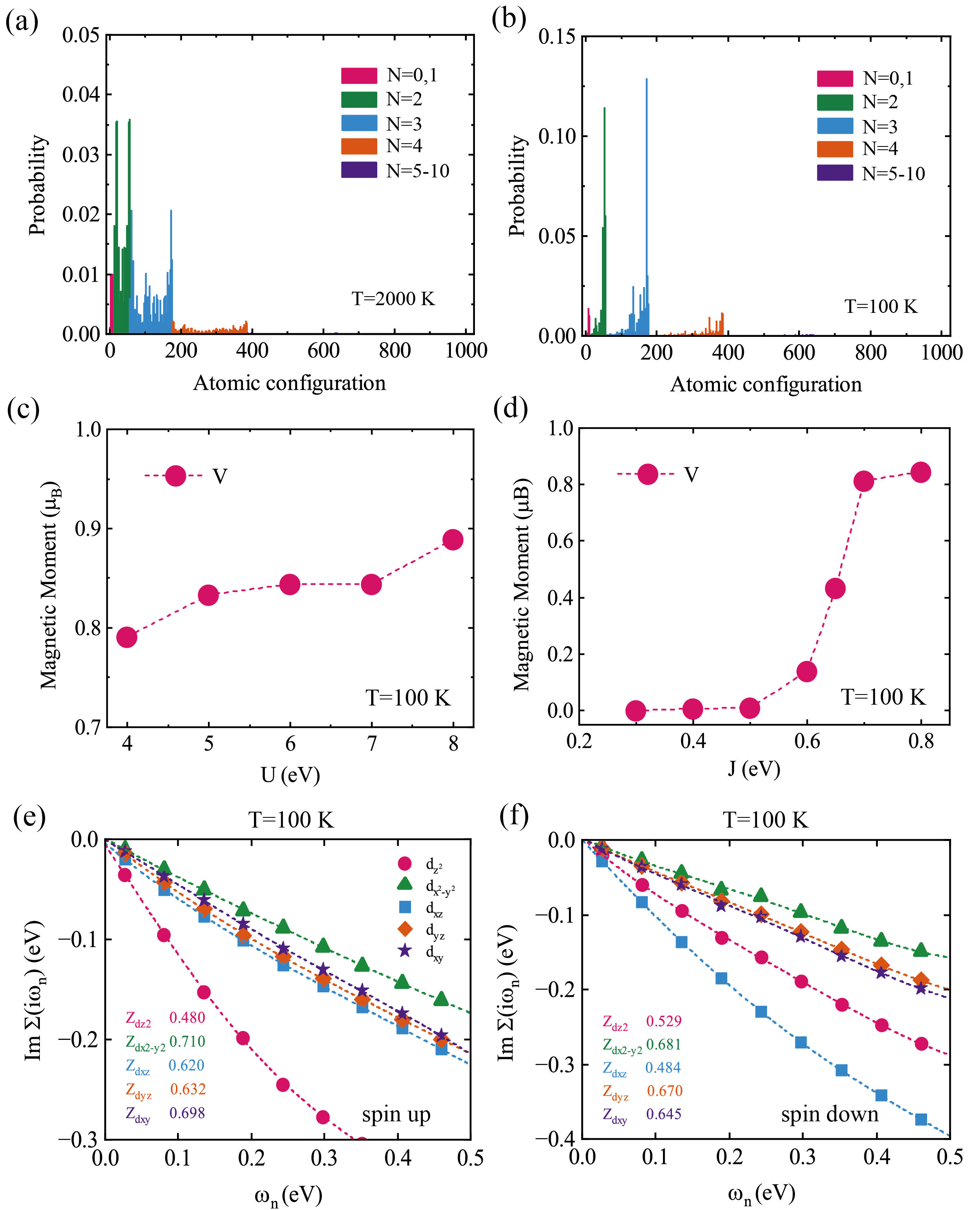}
\caption{Atomic histogram from DFT+DMFT calculations of KV$_{2}$Se$_{2}$O at (a) 2000 K and (b) 100 K. (c) The magnetic moments of V-atoms as a function of Coulomb interaction $U$ with a fixed Hund’s coupling interaction $J$ in DFT+DMFT calculations at 100 K.  (d) The magnetic moments of V-atoms as a function of Hund’s coupling interaction $J$ with a fixed Coulomb interaction $U$ in DFT+DMFT calculations at 100 K. The imaginary part of the orbital-dependent self-energy for V1-atoms at 100 K in DFT+DMFT calculations with spin-up channel (e) and spin-down channel (f). Dotted lines represent fourth-order polynomial fits to the data.}
\label{fig4}
\end{center}
\end{figure}

Previous studies have shown that the DFT-calculated band structures require an upward shift of approximately 90 meV to align with ARPES measurements \cite{Jiang2025}. In particular, the DFT band at the midpoint between the $\Gamma$ and $M$ points exhibits a significantly larger band gap than observed experimentally, as highlighted by the red circle in Fig.~\ref{fig2}(d) \cite{Jiang2025}. These discrepancies suggest that electronic correlation effects may play a critical role in shaping the low-energy electronic structures of KV$_2$Se$_2$O. A comparison between the DFT+DMFT spectral function and the DFT band structure, as shown in Fig.~\ref{fig2}(d), reveals substantial renormalization of the bands near the Fermi level, further supporting the importance of many-body effects. The atomic histograms in Figs.~\ref{fig4}(a) and (b) provide additional insight into the temperature-dependent electronic structures. At 2000 K, the histogram in Fig.\ref{fig4}(a) is nearly symmetric, indicating a lack of magnetic ordering. In contrast, at 100 K in Fig.\ref{fig4}(b), the histogram becomes noticeably asymmetric, reflecting the onset of magnetic order. Moreover, the V atoms exhibit a broad distribution of atomic configurations ($N = 0$ to $N = 4$), a hallmark of Hund’s metal behavior.

Figures~\ref{fig4}(c) and (d) show the evolution of the magnetic moment of the V-atom as a function of the on-site Coulomb interaction $U$ and Hund’s exchange coupling $J$. The results reveal that the magnetic moment is much more sensitive to variations in Hund’s coupling than to changes in $U$. Specifically, as shown in Fig.~\ref{fig4}(c), the magnetic moment remains nearly constant even for $U$ values exceeding 8 eV. In contrast, Fig.~\ref{fig4}(d) illustrates a pronounced enhancement of the magnetic moment as $J$ increases, particularly beyond $J = 0.6$ eV. To further quantify the strength of electronic correlations, we computed the quasiparticle renormalization factor $Z^{-1}=1-\partial \text{Im} \Sigma (i\omega)/\partial \omega |_{\omega \rightarrow 0^{+}}$. As shown in Figs.~\ref{fig4}(e) and (f), the renormalization factor lies in the range of 0.5 to 0.7. These results suggest that KV$_2$Se$_2$O is a mixed-valence system with relatively weak electronic renormalization.

Finally, we revisit the issue of the SDW gap at the Fermi level, as inferred from symmetrized EDCs in ARPES measurements \cite{Jiang2025}. Jiang et al. raised an important question concerning the nature of the SDW gap-like feature, particularly given the experimental challenges inherent to SARPES in altermagnetic systems, such as high noise levels, long acquisition times, and the lack of spectral information above the Fermi level. Notably, the observed pseudogap extends up to 200 K with an estimated size of approximately 0.1 eV, well above the SDW transition temperature of 105 K. By contrast, SDW gaps in iron-based superconductors typically range from 10 to 60 meV \cite{Yi2014}. Furthermore, recent DFT calculations by Yan et al. \cite{Yan2025} indicate that the SDW gap lies significantly below the Fermi level, in agreement with our results. However, the overestimated magnetic moments in standard DFT calculations limit the accuracy in determining the precise position of the SDW gap. These findings suggest that the experimentally reported ``SDW gap” may instead originate from the intrinsic V-shaped density of states characteristic of the altermagnetic phase. More experimental techniques and theoretical methods are required to clarify the true nature of the gap near the Fermi level in KV$_2$Se$_2$O.

In summary, our study identifies KV$_2$Se$_2$O as a metallic altermagnet with robust room-temperature ordering. Dynamical correlation effects lead to reduced magnetic moments, mixed-valence behavior, and band shifts near the $\Gamma$ point. We show that the disappearance of electron-like Fermi pockets near the $\Gamma$ point corresponds to a Lifshitz transition, explaining the observed metal-insulator-metal behavior. The pseudogap at the Fermi level arises from a V-shaped density of states due to orbital- and spin-selective ``half-metal-like'' behavior on V-atoms, distinct from a conventional SDW gap. Given the sensitivity of both the magnetic state and Fermi surface topology, we propose that external pressure could suppress magnetism and induce superconductivity, offering a pathway to study the interplay among altermagnetism, unconventional superconductivity, DW-like instabilities.

{\it  Acknowledgements.} The authors thank Yingying Cao, Min Liu and Yuechao Wang for the fruitful discussions. This work is supported by the National Natural Science Foundation of China (Grants No. 12204033 and No. 52371174) and the Young Elite Scientist Sponsorship Program by BAST (Grant No. BYESS2023301). Numerical computations were performed at the Hefei advanced computing center.

\bibliography{ref}

\end{document}